\documentstyle[preprint,aps]{revtex}
\tightenlines
\input{psfig.sty}
\begin{document}
\preprint{UNDPDK-98-05}
\title{Bounds on $\nu_{\mu}$ Oscillations from Atmospheric Neutrinos}
\author{J.M.~LoSecco}
\address{University of Notre Dame, Notre Dame, Indiana 46556}
\date{\today}
\maketitle
\begin{abstract}

A reanalysis of identified muon neutrino interactions from IMB 3 yields
bounds on $\sin^{2}(2 \theta)$ and $\Delta m^{2}$.  The limit
$\sin^{2}(2 \theta) < 0.72$ is in conflict with the
recent announcement of a neutrino mass.

Subject headings: Cosmic Rays --- Elementary Particles --- Neutrino Oscillations
\\
\end{abstract}


Atmospheric neutrinos have always held the possibility of providing a
large range of propagation coupled with a near source to provide a convenient
calibration.  With a reasonable choice of angular cuts a sample of neutrinos
traveling about 10,000 km can easily be compared with a sample traveling
on the order of 20 km collected in the same experiment at the same time.
Early work \cite{losec85} using these concepts was hampered by low statistics
and poor particle identification.  In \cite{losec85} a mixed sample of 25
upward neutrino interactions was compared with a sample of 25 downward
neutrino interactions.  The neutrino propagation distance is a function of the
neutrino direction, so scattering can, in principle, make the distance
estimate ambiguous.  In reality the propagation distance varies very
slowly with angle near the vertical so as long as regions near the vertical
are compared scattering will have a negligible effect on the sensitivity.
In most analyses regions containing 20\% of the solid angle over the upward
and downward direction have been used.

Depending on the location of the detector the upward and downward fluxes
may be influenced differently by geomagnetic effects\cite{imb88}.  IMB was
located at 52$^{\circ}$ north geomagnetic latitude.  This meant that the local
flux, coming from above, had a lower geomagnetic cut off than the Earth
at large.  As such one would expect a greater rate of downward going events
since the lower cut off permitted more of the extraterrestrial cosmic ray
flux to descend and hit the atmosphere.  But such effects are small and
have not been observed in a statistically significant way even in a sample of
401 neutrino interactions\cite{imb88}.  In the 401 IMB 1 sample 55 events
were measured in the upward going 20\% of the solid angle and 65 were found
in a comparable portion of downward solid angle.  comparison of the 15
upward events and 21 downward events which were classified as muon neutrino
induced because of the presence of a muon decay in the event yielded limits
on $\Delta m^{2}$ in the range of $5 \times 10^{-5}$ eV$^{2}$.

IMB 3 provided a comparable sample of events but with better light collection
and with morphologically based muon identification methods\cite{imb3}
as well as those based on muon decay signatures.
An analysis comparing the shape of the energy distribution for the upward and
downward 20\% of the solid angle\cite{baksan} confirmed the IMB 1 results but
with a larger sample and with particle identification.  The IMB 3 work was
done with 34 downward going and 32 upward going $\nu_{\mu}$.  These two samples
have energy distributions which are statistically indistinguishable.

The excluded region derived in \cite{baksan} is bounded in the range
$ 4.5 \times 10^{-5} < \Delta m^{2} < 1.1 \times 10^{-4} $ eV$^{2}$.  Outside
of this range distortion of the upward going neutrino spectrum by neutrino
oscillations would not make its {\em shape} statistically distinguishable from
the measured downward spectrum.

The contained neutrino interactions can be used to study neutrino oscillations
over a broader range of $\Delta m^{2}$.  The lowest limits on $\Delta m^{2}$
are found in comparisons of the shape of the observed contained event energy
spectra for samples of events near the vertical.  Such a test is insensitive
to detailed flux calculations, since the downward sample constitutes a
measured near source.  The method has a limited range because for larger
$\Delta m^{2}$ the upward going sample would be fully mixed via multiple
oscillations and hence would have a spectrum shape that was similar to the
unoscillated downward sample.  The upward sample would be reduced by an
overall factor, which is $1-\frac{1}{2}\sin^{2}(2 \theta)$.  So comparison of
rate of upward to downward going $\nu_{\mu}$ interactions can extend the
sensitive range.  This comparison eventually fails at even higher
$\Delta m^{2}$ because at large $\Delta m^{2}$ both the upward and downward
going samples will have oscillated.  In such cases oscillations may be
noted by deviations of the observations from expectations.  Estimates
of $\Delta m^{2}$ would be unreliable since no distance scale would be present
in the observations.

The first two methods, comparison of up and down spectral shape and comparison
of up and down interaction rate are insensitive to most systematic errors in
the neutrino flux calculations.  For example they are insensitive to the
normalization uncertainty, the $\nu_{\mu}$ to $\nu_{e}$ ratio uncertainty
and the temporal modulation caused by the solar wind's impact on the Earth's
magnetosphere.  They may be influenced by geomagnetic effects that produce
modest variations of the neutrino flux at various points on Earth.  The
downward neutrino component is produced locally and so is sensitive to local
magnetic field properties.

The result of \cite{baksan} can be extended by comparing the {\em rate}
of upward going to downward going $\nu_{\mu}$ interactions.  For the range
of $\Delta m^{2}$ of greatest interest for the Super Kamioka\cite{suposc}
results one would expect a two to one rate
difference\cite{problem}.

For the data of reference \cite{baksan} the up to down event ratio is
\[
\frac{32}{34} = 0.94 \pm 0.23 > 0.64
\]
where the bound is the 90\% confidence lower limit on the ratio of up to
down flux.  As mentioned above, due to geomagnetic effects this ratio is
expected to be less than one due to a small enhancement of the downward flux.
We neglect the effect of the geomagnetic enhancement of the downward flux
in calculating our neutrino oscillation limits.  This makes the limits a bit
conservative since any reduction in the upward relative rate due to geomagnetic
effects will instead be ascribed to possible oscillation effects.

In the region $2.5\times10^{-4}<\Delta m^{2}<7.0\times10^{-3}$ eV$^{2}$ where
the upward neutrinos have traveled
over several oscillations and the downward neutrinos have not had a chance
to oscillate this bound on the ratio can be converted into a limit
on $\sin^{2}(2 \theta)$.
\[
1-\frac{1}{2}\sin^{2}(2 \theta) > 0.64
\]
\[
\sin^{2}(2 \theta) < 0.72
\]
This is in direct conflict with the results\cite{suposc}
$\sin^{2}(2 \theta) >0.82$ within the mass range
$5\times10^{-4}<\Delta m^{2}<6\times10^{-3}$ eV$^{2}$.
A possible source of this
discrepancy is a misinterpretation of their angular distribution in terms
of low mass neutrino oscillations\cite{problem} by the
Super Kamioka collaboration.

\begin{figure}
\psfig{figure=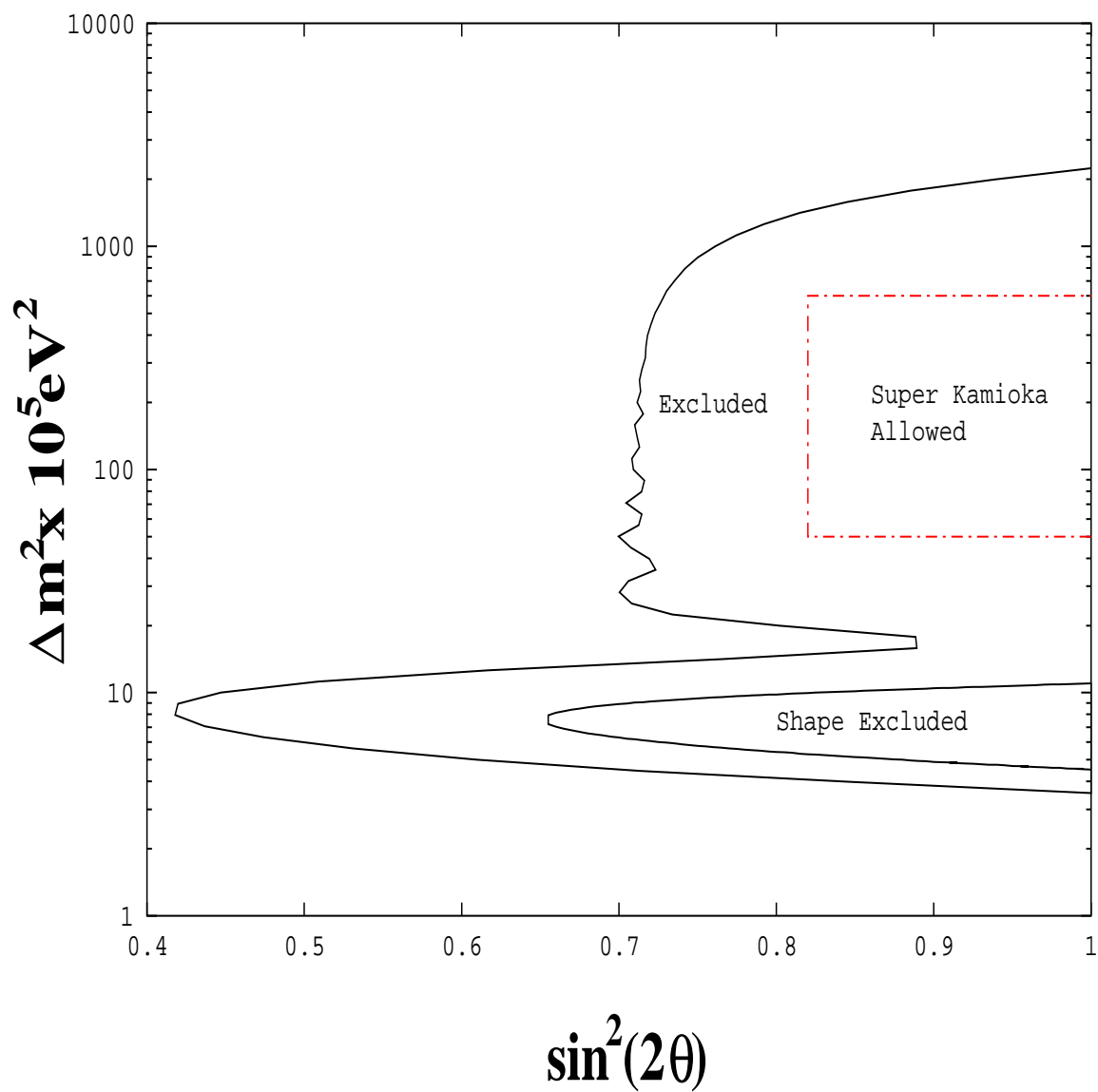,width=6.0in,height=6in}
\caption{\label{excl} 90\% confidence level excluded region}
\end{figure}

A plot of the region excluded by this analysis and the analysis of
reference\cite{baksan} is shown in figure \ref{excl}.  The section marked
``Super Kamioka Allowed'' on the figure is not their true allowed region but
an outline of the maximum range permitted for some values of $\Delta m^{2}$
and some values of $\sin^{2}(2 \theta)$.  The Super Kamioka fit implies
that $\sin^{2}(2 \theta)=1$.   Figure \ref{excl} was calculated by
integrating over the observed energy and distance distribution.

This rate analysis fails to exclude $\Delta m^{2}>2.5 \times 10^{-2}$ eV$^{2}$
because the upper hemisphere
starts to show evidence of oscillations (in this low energy sample) so the
downward rate would also be reduced.  A shape test could be used to extend the
results in this area.  (The fact that the upward rate is not significantly
{\em greater} than the downward rate permits one to exclude the
range $\Delta m^{2} < 0.1$ eV$^{2}$.  To calculate detailed limits one must
integrate over the source to detector distance since the muon decay length
is comparable to the path length.)
The region $6.3 \times 10^{-5}<\Delta m^{2}<1.0 \times 10^{-4}$ is more
sensitive to rate effects than our simple analytical
analysis above indicated.  In this mass region the oscillation hypothesis takes
the bulk of the upward data through the maximum.  The ripples in the figure are
indications of the repeat of this effect for the second and third
oscillation {\em etc.}

These results confirm the excluded region calculated from a study of an
independent sample of upward going muons by IMB\cite{nuosc}.

A larger sample of IMB 3 contained event data\cite{imb3} is available and
could be used to extend these results.  With double the data sample the upper
bound on $\sin^{2}(2 \theta)$ would drop to 0.54.  The Super Kamioka single
ring showering events \cite{suposc} should yield a limit on
$\nu_{e} \rightarrow \nu_{e}$
of the order of $\sin^{2}(2 \theta_{ex}) < 0.23$ over a comparable range of
$\Delta m^{2}$.  Precise contours depend on details of the {\em measured}
$\nu_{e}$ energy distribution.

\section*{Acknowledgments}
I am grateful to the The Institute for Nuclear Research of the USSR Academy
of Sciences for the opportunity to participate in their Particles and
Cosmology School in 1991.  I am grateful to Richard Feynman for suggesting
we extend the results of reference \cite{losec85} to
higher $\Delta m^{2}$.

\end{document}